\documentclass{aa}
\usepackage{graphicx}
\begin{document}
   \title{First microlensing candidate towards M31 from the Nainital Microlensing
Survey}
     
   \author{Y. C. Joshi\inst{1,2},  A. K. Pandey\inst{2}, D. Narasimha\inst{1,3},
R. Sagar\inst{2}}

   \offprints{Y. C. Joshi ~(ycjoshi@tifr.res.in)}

    \institute {Tata Institute of Fundamental Research, Homi Bhabha Road,
Mumbai -- 400 005, India
\and
Aryabhatta Research Institute of Observational Sciences (ARIES), Manora peak
Nainital - 263129, India 
\and
Astronomical Institute, Tohoku University, Sendai 980-8578
}

   \date{Received 12 November 2004 / Accepted 20 December 2004}

\abstract{
We report our first microlensing candidate NMS-E1 towards M31 from the data
accumulated during the four years of Nainital Microlensing Survey. Cousin $R$
and $I$ band observations of $\sim 13^{'} \times13^{'}$ field in the direction
of M31 have been carried out since 1998 and data is analysed using the pixel
technique proposed by the AGAPE collaboration. NMS-E1 lies in the disk of M31
at $\alpha = 0^h 43^m 33^s.3$ and $\delta = +41^{\circ} 06^{'} 44^{''}$, about
15.5 arcmin to the South-East direction of the center of M31. The degenerate
Paczy\'{n}ski fit gives a half intensity duration of $\sim$ 59 days. The
photometric analysis of the candidate shows that it reached $R \sim 20.1$ mag
at the time of maximum brightness and the colour of the source star was
estimated to be $(R-I)_0 \sim 1.1$ mag. The microlensing candidate is blended
by red variable stars; consequently the light curves do not strictly follow
the characteristic Paczy\'{n}ski shape and achromatic nature. However its long
period monitoring and similar behaviour in $R$ and $I$ bands supports its
microlensing nature.
}
\authorrunning{Y. C. Joshi et al.}
\titlerunning{First microlensing candidate towards M31 from the Nainital
Microlensing Survey}
\maketitle
\section{Introduction}
Recent years have seen rapid advances in our understanding of the nature of
dark matter in the Galaxy. Much of this progress has come from the studies of
gravitational microlensing, a phenomenon of temporary amplification of the
flux of a background star when a foreground MACHO passes close to our line of
sight to this background star. In the last decade, several collaborations have
focused their attention on Galactic bulge and Magellanic Clouds following
the suggestion of Paczy\'{n}ski (1986) and reported more than one thousand
microlensing events after monitoring millions of star in these directions
(e.g. Alcock et al.~1993 for MACHO, Udalski et al.~1993 for OGLE and Aubourg
et al.~1993 for EROS collaborations). These groups have ruled out most of the
dark matter in the form of sub-stellar objects and estimated that only up to
20\% of the halo could be made up of MACHOs in the mass range of $0.15M_\odot$
to $0.9M_\odot$ (Alcock et al.~2000., Lasserre et al.~2000).

Although the conventional microlens monitoring is successful towards the
Galactic bulge and Magellanic Clouds, it is fundamentally limited to only
nearby regions where there are large number of resolved stars. In regions
like M31, which is at $\sim$ 780 kpc (Stanek \& Garnovich 1998, Joshi et
al.~2003), microlens monitoring is a cumbersome process due to the large
background of unresolved stars of M31 where the typical stellar density is as
high as few hundred stars per arcsec$^2$. To deal with this problem, Baillon
et al.~(1993) proposed to monitor pixels of the CCD rather than the stars
themselves. Any variation in the flux of stars due to microlensing would be
determined in the corresponding variation in the pixel of the CCD. Therefore,
by monitoring a pixel for a long time, one could detect a microlensing event.
The pixel technique, which relies on the shape analysis of the light curve to
detect microlensing events, is implemented by the AGAPE (Ansari et al.~1997)
and POINT-AGAPE collaborations (Paulin-Henriksson et al.~2002) and have
reported more than 10 possible microlensing events towards M31 (Calchi Novati
et al.~2002, 2003, Paulin-Henriksson et al.~2002, 2003). A similar approach
but based on the image subtraction technique was proposed by Crotts (1992) and
Tomaney \& Crotts (1996) that has been further improved by Alard \& Lupton
(1998) and Alard (2000) by optimal image subtraction using a space-varying
kernel. The technique is widely used to detect different kinds of variable
sources (cf. Bonanos et al.~2003; Taylor 2004). Recently the MEGA collaboration
reported 14 microlensing candidates in their survey towards M31 using the
image subtraction technique (de Jong et al.~2003). Two of them (ML-7 and
ML-11) had already been reported by POINT-AGAPE in their survey using the
pixel technique (PA-99-N2 and PA-00-S4 respectively). Both the groups give
similar results although derived by two different techniques.

Although pixel lensing technique allows us to detect microlensing events where
the standard technique of microlens detection is not applicable, it is limited
to only bright stars that are either giants and super-giants or stars that are
amplified enough to become detectable above the noise level of background flux.
Since the search for gravitational microlensing events requires monitoring of
millions of stars over a few years to yield a significant number of detections
as well as to test their uniqueness, we started multi-band observations of M31
in the autumn of 1998 under a program {\bf Nainital Microlensing Survey} at
the ARIES, Nainital, India in collaboration with the AGAPE (Joshi 2004). The
details of our observations are given in the next section while the pixel
technique and its implementation is described in Sect. 3. The selection
criteria for the microlensing search and our results are given in Sect. 4
followed by the discussion and conclusions.
\begin{figure}
\includegraphics[width=9.0cm, height=8.8cm,angle=270]{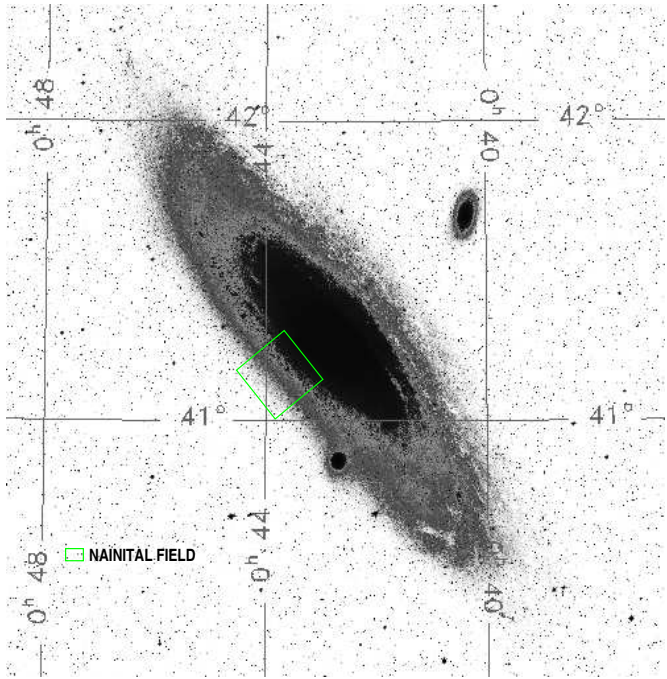}
\vspace{0.5cm}
\caption{A $13'\times 13'$ field of Nainital Microlensing Survey juxtaposed
over M31.}
\end{figure}
\section{Observations}
We started Cousins $R$ and $I$ band observations in the direction of M31 in
1998, with a small 1K$\times$1K CCD covering an area of $\sim 6'\times 6'$ at
the f/13 Cassegrain focus of 104-cm Sampurnanand Telescope. Since 1999,
observations have been carried out with a larger size 2K$\times$2K CCD chip
which covers $\sim 13'\times 13'$ encompassing the previous field. Each pixel
covers an area of 0.37$\times$0.37 arcsec$^2$ of the sky. The target field is
centered at $\alpha _{2000}$ = $0^{h} 43^{m} 38^{s}$ and $\delta_{2000}$ =
$+41^{\circ}09^{\prime}.1$, at a distance of about 15 arcmin in the
South-East direction from the center of M31. Fig.~1 shows the location of
the target field superposed over the image of M31. In the 4 year observing
run, we accumulated a total of 468 data points in $R$ band and 383 data points
in $I$ band during 141 observational nights spanning $\sim$ 1200 days. A
histogram of the data taken in the survey is shown in Fig.~2. The average
seeing during the observations was $\sim$ 2.2 arcsec. A detailed overview of
the observations is given in Joshi et al.~(2003).
\begin{figure}
\includegraphics[width=9.0cm, height=11.0cm]{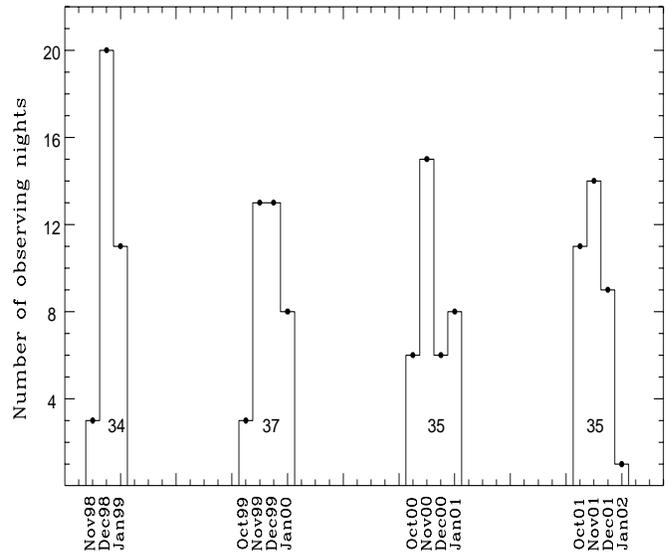}
\vspace{-3.5cm}
\caption{A histogram of the data taken in the 4 year survey.
The total number of nights observed per observing season is given in the box.}
\end{figure}
\section {Pixel method}
The basic principle of the pixel method is that if a star of flux $F_*$ is
lensed in a crowded field, then by subtracting the original flux from the
amplified flux of the star, one gets an increase in flux equal to $(A-1)F_*$
above the photon noise where $A$ is the magnification of star. Now if the
target star shows a variation in the flux with time, it reflects in the
$\Delta F$ of the pixel. Thus by following $\Delta F$ with time, we can
monitor the variation in the flux of the target star. Although to detect any
variation in the flux, this change should be significantly above
the fluctuation level.  

In the point-like lens and uniform motion approximation, the magnification
factor $A$ is related to the impact parameter $u$ as given by the following Eq.
\begin{equation}
A(t) = \frac{u^2(t)+2}{u(t)\sqrt{u^2(t)+4}}
\end{equation}
where $u^2(t) = u^2_0 + \frac{(t-t_0)^2}{t_E^2}$. Here $t_0$ and $t_E$ are the
time of maximum amplification and Einstein time scale respectively and $u_0$
indicates the distance of closest approach in the units of Einstein radius.
When $u(t) << 1$, the magnification is well approximated by
$$A(t) \equiv \frac{1}{u(t)}; ~~~~ A_{max} \approx \frac{1}{u_0}$$
Then the change in flux of the target star is given by
\begin{equation}
\Delta F = \frac{F_* A_{max}}{\sqrt {1 + \biggl(\frac{t-t_0}{t_E/A_{max}}\biggr)^2}}
\end{equation}
From the light curve of a resolved star, one can get three parameters, namely
peak time $(t_0)$, Einstein time $(t_E)$ and peak magnification $(A_{max})$
by fitting a theoretical light curve. However, for the unresolved stars as in
the case of pixel lensing, there is a degeneracy between the $t_E$, $A_{max}$
and $F_*$ (Gould 1996). Hence, high quality data (e.g. Paulin-Henriksson et
al. 2003) or detection of the source star (e.g. Auri\'{e}re et al.~2001) are
required to accurately determine the $t_E$ and hence the mass of the MACHO.
\subsection {Implementation of the pixel method in Nainital data}
Standard CCD reduction technique were used for the pre-processing of data
using IRAF\footnote{Image Reduction and Analysis Facility (IRAF) is
distributed by the National Optical Astronomy Observations} which includes
bias subtraction, flat fielding and cosmic ray removal. We added all the
frames of a particular filter taken on a single night and made one frame per
filter per night to increase the signal to noise ratio.
\subsubsection {Alignments}
To align the images, we have to first select a reference frame taken in good
photometric condition with lower sky background and relatively small seeing.
We chose an image taken on January 05, 2000 as the reference frame. The
average seeing during the night was 1.5 arcsec. 

Since the observations were carried out on different nights, a star does not
fall onto the same pixel in all the images. Therefore, we align all the images
through rotation and shifting with respect to the reference frame. We achieve
an accuracy better than 0.05 arcsec in alignment for most of the images.
\begin{figure}[h]
\centering
\includegraphics[width=8.0cm, height=8.0cm]{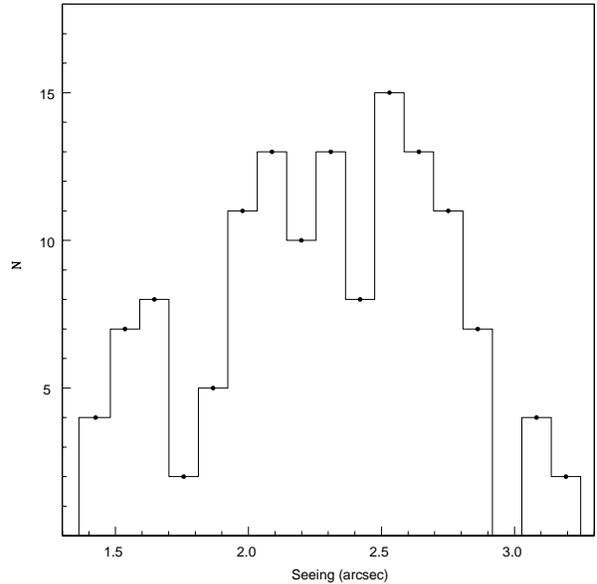}
\caption{Seeing distribution during the four year observing period.}
\end{figure}
The photometric conditions like the Earth's atmospheric absorption and sky
background are never the same for different nights. To bring all the images
at same photometric level, we correct the images with respect to the reference
frame using the statistical approach explained by Ansari et al.~(1997). To
overcome the residual gradient between two different frames due to reflected
light, we normalize all the images with respect to the median background of
the reference frame. 
\subsubsection {Seeing correction}
Most of our observations were carried out in the early hours of the night when
sky conditions might not have been perfectly stable, resulting in large seeing
which is conspicuous in Fig.~3 where we have plotted the histogram of seeing
variation with time in the four year observing run. The seeing during our
observations varies in the range $\sim 1^{''}$.4 to 3$^{''}$.2 with an
average seeing of 2$^{''}$.2. The large seeing fluctuations during different
observing periods may cause a false variation in the pixel light curve. To
correct it, we make a superpixel of $7\times7$ pixel$^2$ ($\sim 2.5\times2.5$
arcsec$^2$ area of sky) around each pixel so that most of the stellar flux
falls on that superpixel\footnote {In the subsequent sections, we shall use
the term `pixel' instead of `superpixel'} although it only solves part of the
seeing problem and does not correct
\begin{figure}[h]
\centering
\includegraphics[width=9.0cm, height=12.5cm]{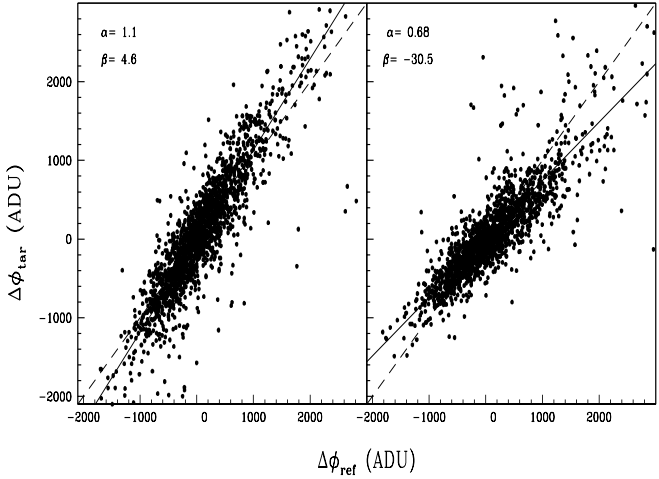}
\vspace{-6.3cm}
\caption{Linear correlation between the quantities
$\Delta\phi_{tar} \equiv (\phi_{tar}-\tilde{\phi}_{tar})$ and
$\Delta\phi_{ref} \equiv (\phi_{ref}-\tilde{\phi}_{ref})$.
The left and right graphs show the correlation for the images of good
$(1^{''}.4)$ and bad $(2^{''}.9)$ seeing. The seeing of the reference image is
1$^{''}$.5. The dashed lines is the y = x line while the continuous lines
indicate y = $\alpha$ x+$\beta$ line.}
\end{figure}
\begin{figure}[h]
\centering
\includegraphics[width=8.5cm,height=10.5cm]{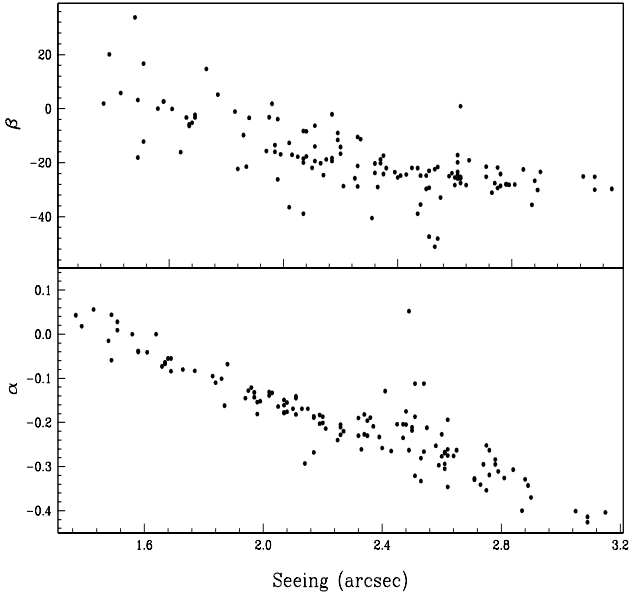}
\vspace{-2.3cm} 
\caption{The variation of seeing correction coefficients $\alpha$ and
$\beta$ as a function of seeing.}
\end{figure}
for the loss of photon counts in the changing wings of the PSF due to
highly variable seeing between the frames. We thus correct the images by
using the AGAPE `seeing stabilization method' (Calchi Novati et al.~2002,
Paulin-Henriksson et al.~2002) where we assume on the basis of empirical
observations that the difference in flux between any image and its median
background is linearly correlated to the same difference for the reference
image i.e.
\begin{equation}
\phi_{i} - \tilde{\phi}_{i} = \alpha_{i} (\phi_{ref}-\tilde{\phi}_{ref}) + \beta_{i}
\end{equation}
where $\phi_{ref}$ and $\phi_{i}$ are the flux of reference and $i^{th}$ images
respectively while $\tilde{\phi}_{ref}$ and $\tilde{\phi}_{i}$ are the flux of
their respective median background images. The $\alpha_{i}$ and $\beta_{i}$
are the seeing correction coefficients for the $i^{th}$ image. In Fig.~4, we
show the linear correlation between $(\phi_{ref}-\tilde{\phi}_{ref})$ and
$(\phi_{curr} - \tilde{\phi}_{ref})$ for two images, one taken in good seeing
and other one in bad seeing conditions. The $\alpha_{i}$ and $\beta_{i}$,
which are calculated with a $\chi^2$ minimization procedure, shows a
correlation with the seeing as shown in Fig.~5. After determining $\alpha$ and
$\beta$ for each image, we correct them by
\begin{equation}
\phi_{i}^{new} =  \frac{\phi_{i} - \tilde{\phi}_{i} - \beta_i}{\alpha_i} + \tilde{\phi}_{ref}
\end{equation}
where $\phi_{i}^{new}$ is the modified flux after seeing correction. This
correction scales all the images to the seeing of reference image.
\subsection {Pixel light curves}
To derive a pixel light curve, we mask bad pixels which are either bright
stars for which photometric analysis has already been carried out or have
contaminated due to spill-over of the saturated stars. In this way, we reject
about 20\% of the total pixels from the pixel analysis.
\subsubsection {Stability of light curves}
Since all the images are geometrically and photometrically aligned as
well as seeing corrected, we expect the pixel light curve of non-variable
objects to be stable with time. In Fig.~6, we illustrate the light curves
of a stable pixel (782,579) in both $R$ and $I$ bands during its three
year monitoring. The R.M.S. fluctuations along the light curve are $\sim$ 97
and $\sim$ 108 ADUs in $R$ and $I$ bands respectively which are less than
the average error bars of $\sim$ 166 and $\sim$ 292 ADUs in the respective
bands. This demonstrates that we have achieved a high level of stability in
the pixel light curves after applying all the corrections.          
\begin{figure}[h]
\centering
\includegraphics[width=9.0cm,height=9.0cm]{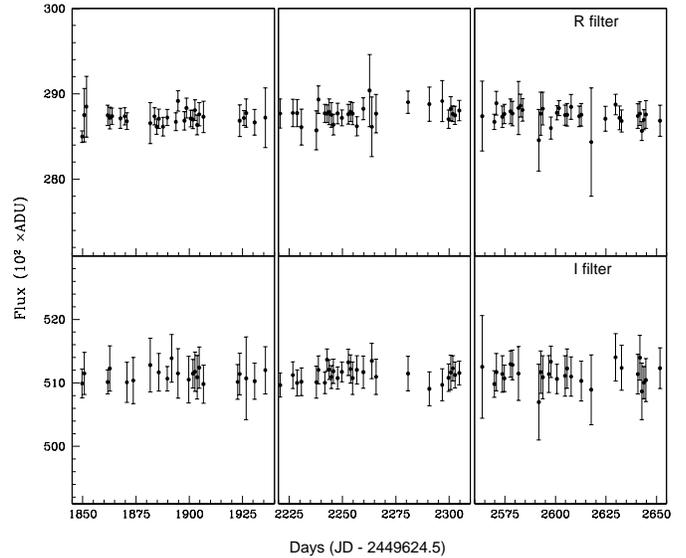}
\vspace{-2.0cm}
\caption{The light curves of a non-varying pixel (782,579) in $R$ and $I$
bands during the 1999-2001 observing season.}
\end{figure}
\subsubsection {Comparison: pixel vs photometric light curves}
To check the robustness of the pixel technique for our data, we compare the
pixel light curve of a star with its complementary light curve derived
through the photometric technique. However, for the comparison, we cannot
take a very bright star, where the seeing correction used in the pixel
technique does not work well nor a very faint star where the photometric error
in magnitude determination is too large. Therefore, we chose a Cepheid
of $R = 20.5$ mag having pixel coordinates (1186,595) for the
comparison.
\begin{figure}[ht]
\label{stable}
\centering
\includegraphics[width=9.0cm, height=12.0cm]{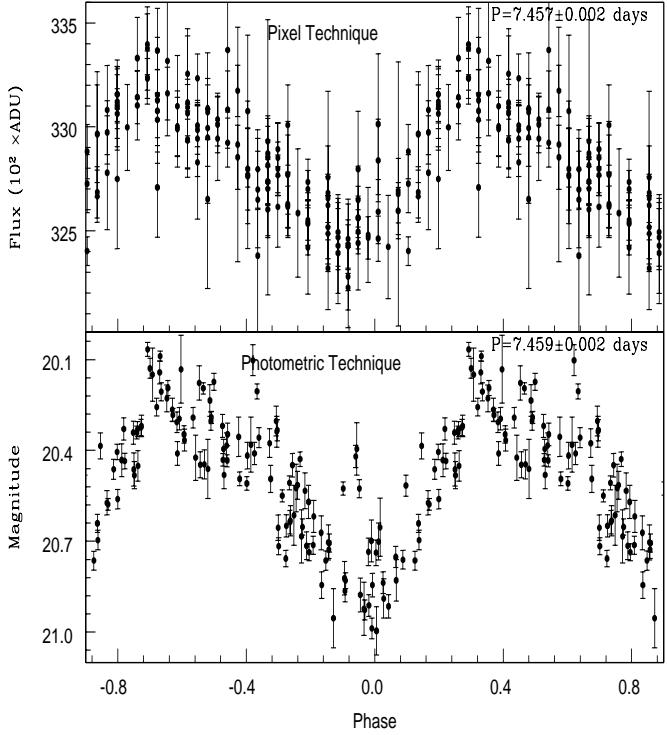}
\vspace{-2.4cm}
\caption{The $R$ band phase light curve of Cepheid (1186,595) derived with
pixel and photometric techniques. Phase is plotted twice and in such a way
that the minimum brightness falls at zero phase. The flux and the magnitude
derived are not scaled.}
\end{figure}
From the data derived with pixel technique, we obtain a period of
7.457$\pm$0.002 days for the Cepheid using the period-searching program by
Press \& Rybici (1989). This value is in agreement with the period of
7.459$\pm$0.002 days determined by Joshi et al.~(2003). This supports the
robustness of the pixel technique implemented on our data. For a comparison,
the phase light curves obtained by these two different techniques are shown
in Fig.~7. Nevertheless, it is seen that the photometric technique is better
for the resolved/bright stars while the pixel technique is ideally suited in
the case of unresolved/faint stars.
\section {Search for microlensing events}
The pixel method relies on the monitoring of the flux variation of a pixel with
time. To identify candidate pixels for the microlensing event, we adopt the
same procedure as defined by Calchi Novati et al.~(2002). Since $R$ band
data have a better sampling density and photometric quality than that of the
$I$ band data, we used only $R$ band data to select the light curves that are
compatible with the microlensing event and $I$ band data will be used to
analyse the microlensing candidate pixels. In the following sub-sections we
briefly summarize the adopted procedure.
\subsection {Selection criteria}
Initially we define a base level which is the minimum value of the sliding
average flux of 9 consecutive points on the light curve. We thereafter
define a bump in the light curve as 3 consecutive points rise above the 3 sigma
level and 2 consecutive points fall below the 3 sigma level. Since there is
about a nine month gap between two consecutive observing seasons, we
constrain all these points to be in the same observing run. For each
bump, we define a likelihood function as
\begin{equation}
L = -ln\Big(\prod_{i \in bump} P(\phi > \phi_i | < \phi >_{i,\sigma_i})\Big)
\end{equation}
where
\[
P(\phi|\phi>\phi_i) = \int_{\phi_{i}}^{\infty}\frac{1}{\sigma_i
\sqrt{2\pi}}exp\big[-\frac{(\phi-\phi_{bkg})^2}{2\sigma_i^2}\big]d\phi
\]                                                                      
We chose only those pixels that have a primary bump with $L > 300$ and no
secondary bump with $L>70$ in the light curve. To exclude all those light
curves that have more than one significant variation, we limit the
ratio $\frac{L_{secondary}}{L_{primary}}$ to be not greater than 0.1. 
We find $\sim$ 19000 pixels that satisfy the above selection criteria.
Most of them are clusters of pixels associated with significant physical
variation. We cluster these pixels and select only those pixels
that have the highest value of $L$ in their immediate neighbouring pixels.
In this way we find 912 pixels characterized by the mono bump. To further
check the flatness of the baseline of these pixels, we determine $\chi^2/N$
taking only those points which are used to determine baseline. A histogram
of the $\chi^2/N$ for these pixels are shown in Fig.~8. To select the bumps
\begin{figure}[h]
\label{baseline}
\centering
\includegraphics[width=9.0 cm, height=8.0 cm]{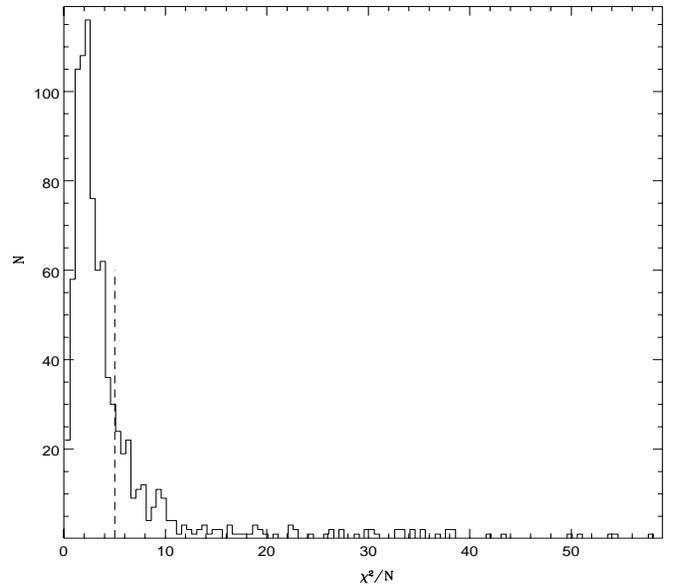}
\caption{$\chi^2/N$ for the baseline flux. A vertical dashed
line is drawn at $\chi^2/N = 5$, above which all the pixels were rejected.}
\end{figure}
with a flat baseline, we reject all those pixels that have $\chi^2/N > 5$.
This selection reduces the pixels to only 670 that are used for the shape
analysis test.           
\begin{figure*}
\centering
\vspace{0.3cm}
\includegraphics[width=8.0cm, height=7.0cm]{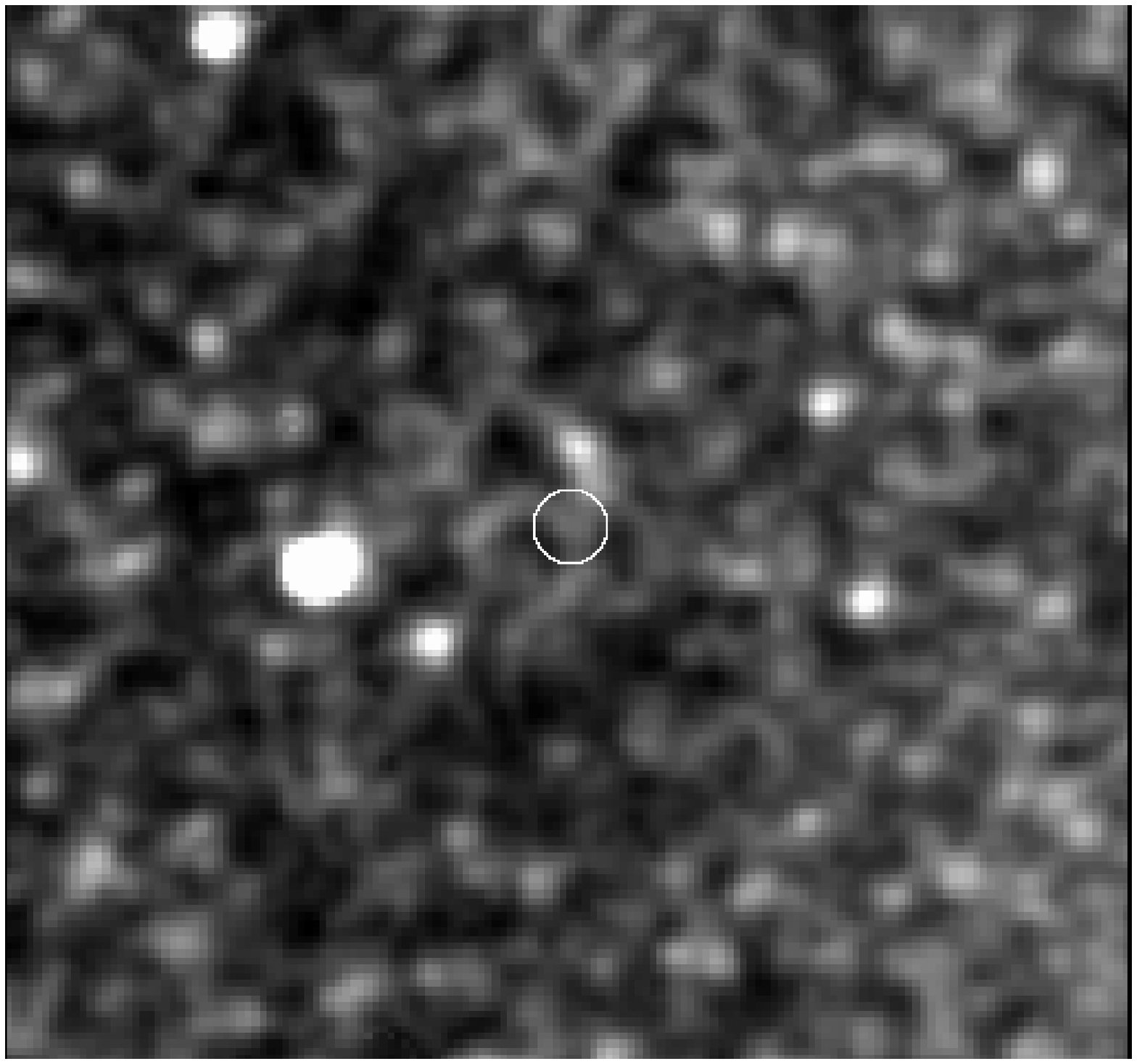}
\hspace{-0.8cm}
\includegraphics[width=8.0cm, height=7.0cm]{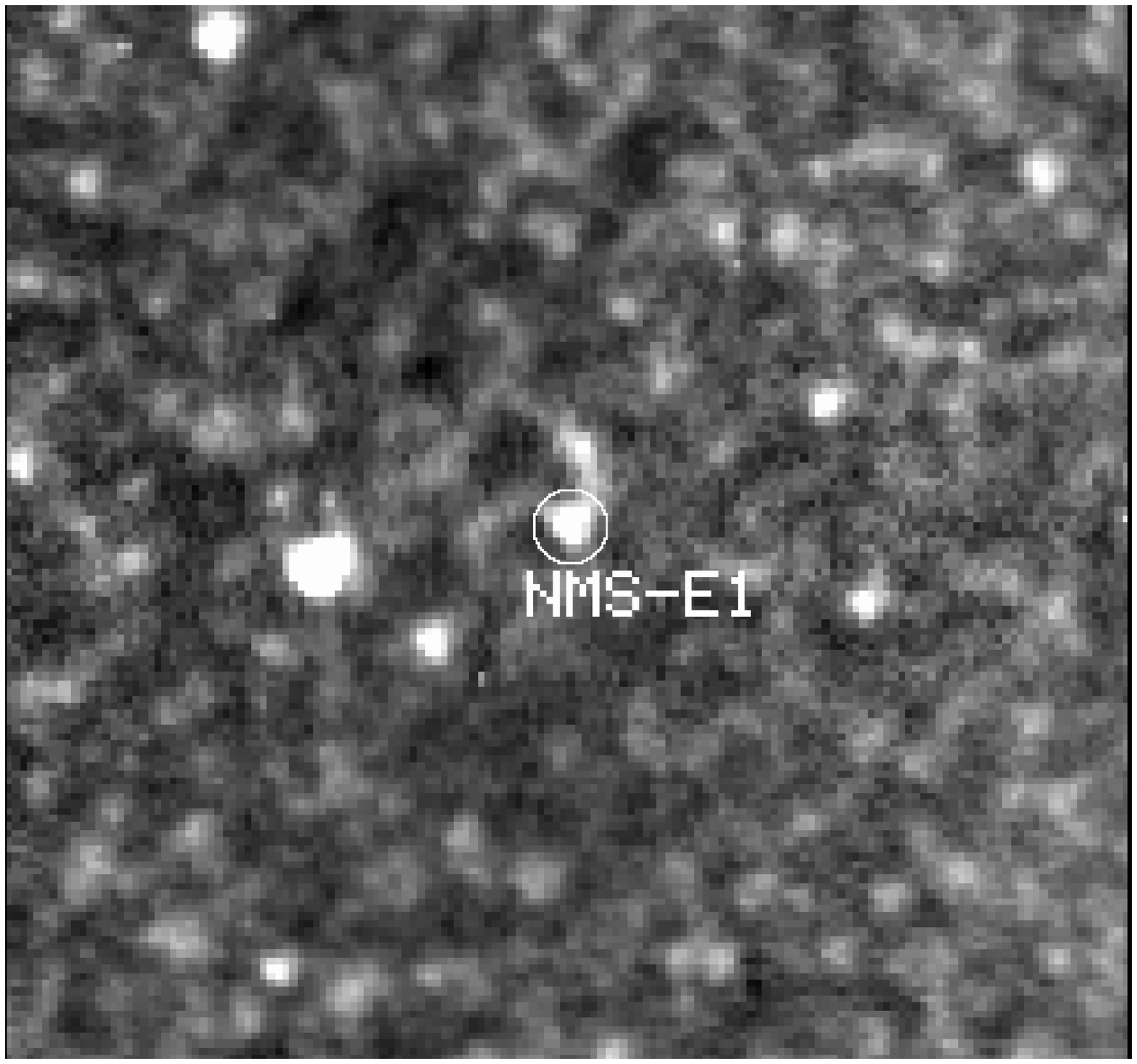}
\caption{The position of NMS-E1 shown by the circle. The images are taken in
$I$ band. The left frame shows no brightness at pixel (880,1402) while
the right frame shows a resolved star at that position.}
\end{figure*}
We do not enforce any selection criteria on $t_{1/2}$ as performed in some
other surveys (e.g. Paulin-Henriksson et al.~2003, Riffeser et al.~2003)
because of: $a)$ observational gaps in our data of up to 10 days between two
consecutive nights within a single observing season; $b)$ the duration of the
observations per season with an average observing density of once in three
days; $c)$ the total period of our monitoring ($> 800$ days)that is large
enough to exclude any other kind of variability and $d)$ we are observing
towards the outer disk of M31 where one generally expects long time scale
events. Since the concentration of variable stars having longer period is
higher as well as blending due to being greater seeing, microlensing
candidates suffer from inhomogeneity in the light curve. We therefore
explicitly check each light curve by eye and eliminates all those pixels
having a periodic variability following Paulin-Henriksson et al.~(2003). We
further fit a 5-parameter Paczy\'{n}ski fit on the $R$ band data and a
non-linear least squares fit estimated as
\[
\chi^2 =  \sum_{i=1}^{n}\frac{\big(\phi(t_i)-\phi_{fit}(t_i)\big)^2}{\sigma^2(\phi(t_i))}
\]
and determine the $\chi^2$ per degree of freedom ($\overline\chi^2$). We
reject those light curves that has $\overline\chi^2 > 4$. This left us with
7 pixels. To further check the achromatism and symmetry in the selected
pixels, we now use both $R$ and $I$ band data together. We find that 5 of
them show significant variation in $I$ band while one of them shows no
\begin{table}[h]
\centering
\caption{Selection criteria to determine microlensing candidates.}
\begin{tabular}{l|l} \hline
Condition & pixels left\\ \hline
Masking bad pixels & 80\% \\
Bump detection & $\sim$19000 \\
Clustering of pixels & 912 \\
$(\chi^2/N)_{baseline} < 5$  & 670 \\
$\overline\chi^2 < 4$ & 7\\
7-parameter fit & 1\\
\hline
\end{tabular}
\end{table}
synchronized variation in $R$ and $I$ bands
as well as a large value of its half intensity duration hence needs further
monitoring. This finally gives us one pixel which we have taken as a genuine
microlensing candidate named as NMS-E1 where NMS is an acronym of the
`Nainital Microlensing Survey'. A summary of the selection criteria is
given in Table 1.
\subsection {Microlensing Candidate NMS-E1}
We find microlensing candidate NMS-E1 at the pixel coordinate (880,1402)
corresponding to $\alpha_{2000}=00^{h}43^{m}33.3^{s}, \delta_{2000}=
+41^{\circ}06^{'}44^{''}$ with an error of $\sim 1^{''}$ in both $\alpha$ and
$\delta$. The candidate lies at an angular distance of $15^{'}28^{''}$ on the
far side of the disk from the M31 center and its position in our frames is
shown in Fig.~9. The source is unresolved at its minimum brightness in our
observations as can be seen to the left side of the figure, however, it is
clearly visible when it reached its maximum brightness which is shown to the
right side of the figure. The $R$ and $I$ band pixel light curves of NMS-E1
are shown in Fig.~10. A 7-parameter (i.e. full width at half maximum
$t_{1/2}$, time of maximum amplification $t_0$, impact parameter $u_0$
and baseline flux and stellar flux for both the bands) Paczy\'{n}ski fit to
the pixel light curves gives a half duration time of $t_{1/2} \sim 59\pm 2$
days with a peak magnification time of $t_0 = 1908\pm1$ days. 
\begin{figure*}
\centering
\includegraphics[width=17.0cm, height=11.5cm]{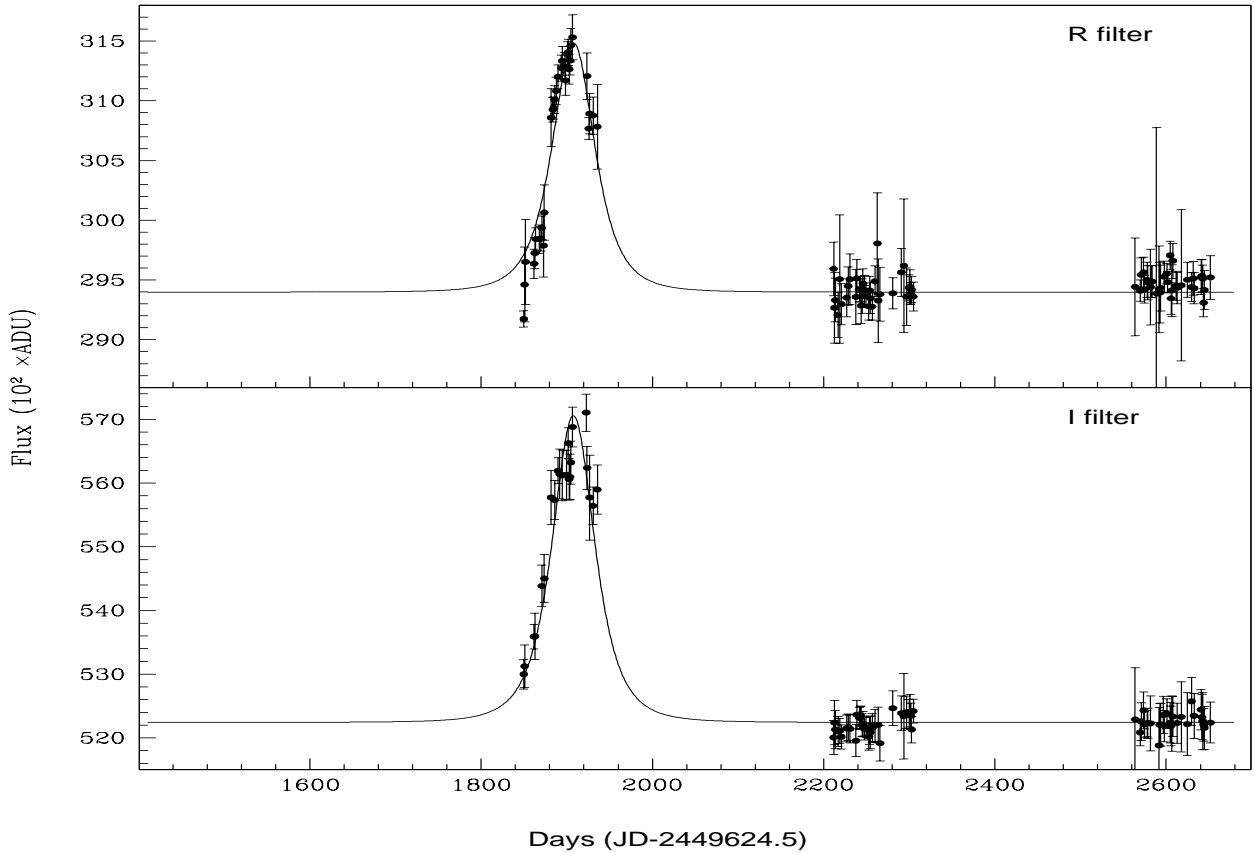}
\caption {The $R$ and $I$ band pixel light curves of NMS-E1 in the upper and
lower panels respectively. The continuous lines represent the results of the
7-parameter Paczy\'{n}ski fit.}
\end{figure*}
\subsection{Shape analysis and achromaticity}
A microlensing event possesses a characteristic Paczy\'{n}ski shape
(Paczy\'{n}ski 1986) with a flat baseline and achromatic nature. The event
NMS-E1 does not strictly follow symmetric shape as is evident from Fig.~10. 
It is seen that the shape of the light curve is broadened in $I$ band in
comparison to the $R$ band.

When we have seen pixels around (880,1402), we find a bright source at
$\sim 2^{''}.5$ to $\sim 5^{''}.5$ south-west that reach up to $R \sim$ 20.6
magnitude having a mean $(R-I) \sim 1.3$. Its photometric analysis indicates
that the source could be a red variable. We did not find any other bright star
in our observations at the detection limit of $R \sim 21.5$ mag. When we
analyse the light curve of the same pixel in MDM data\footnote {The pixel
coordinate corresponding to NMS-E1 in the MDM target field is (1227,451). The
light curve for this pixel was kindly provided by the AGAPE collaboration.}
which is observed with the 1.3 m telescope and better seeing conditions, we
see a bump in the light curve during the 1998-1999 observing season along with
a few overlapping points in the rising branch of the microlensing candidate.
We estimate the magnitude of the bump equivalent to
\[
R(\Delta \phi) \sim 22.8; ~~~~ I(\Delta \phi) \sim 20.7
\]
This star seems very close to the candidate pixel and cannot be resolved
in our images. Although we cannot exactly estimate the magnitude of the star,
its change in flux in the MDM light curve indicates that it is an extremely
red variable. As each pixel of $\sim 2^{''}.5 \times 2^{''}.5$ contains
more than one thousand stars of M31, it is very likely that the nearby bright
stars and red variables contaminate the pixel flux which not only deforms the
shape of light curve but may also change the achromatic nature of the event.

\begin{figure}
\centering
\includegraphics[width=9.0cm, height=9.0cm]{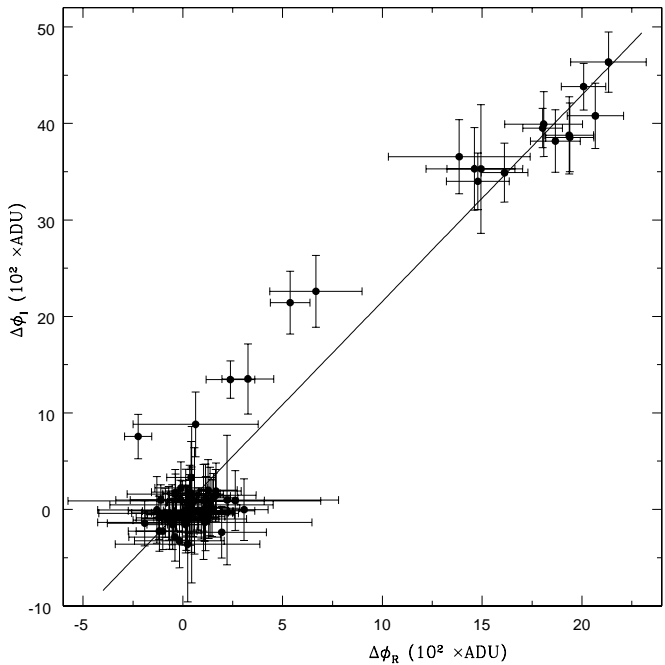}
\caption{Flux-flux variation of NMS-E1 where the x- and y-axis show the
increase of flux in $R$ and $I$ bands respectively. The points around (0,0)
indicate the base of the light curve. A continuous line indicates the best fit
linear regression line.}
\end{figure}
To test the achromatic nature and effect of blending on the candidate
event NMS-E1, we have drawn a flux-flux diagram in Fig.~11. The fit
($\chi^2/ndf = 1.6$) is linear for the upper portion of the light curve within
our noise level but it shows a systematic increase of flux in $\Delta(\phi_I)$
around the fainter side of the microlensing regime which suggests
that the light curve of NMS-E1, in general, is chromatic. However, keeping in
mind the nearby red variable star contamination as we discussed earlier, it
is not unexpected. In crowded regions like M31, it is quite difficult to
separate out the noisy contribution from nearby variable sources. The effect of
blending becomes more severe in the I band where the amplitude of variability is
generally higher. The large and variable seeing in our observations further
worsens the contamination problem. Most of the pixel lensing candidates
towards M31 are severely contaminated by the nearby red variable stars which
is also reflected in the various candidates reported by Calchi Novati et
al.~(2002, 2003), Paulin-Henriksson et al.~(2002) and de Jong et al.~(2003). 

The physical parameters of NMS-E1 derived with the pixel technique as well
its photometric analysis are summarized in Table 2.
\begin{table}[h]
\centering
\caption{Characteristics of the event NMS-E1.}
\begin{tabular}{l|l} \hline
$\alpha$ (J2000)& $00^{h}43^{m}33.3^{s}$   \\
$\delta$ (J2000)& $+41^{\circ}06^{'}44^{''}$ \\
distance from M31 center & $15^{'} 28^{''}$\\
$t_{0}$ & 1908$\pm$1 \\
$t_{1/2}$& 59$\pm$2 days \\
$R_{max}$ & 20.1 mag \\
$(R-I)$ & 1.3$\pm$0.2 mag \\
$\chi^2$/d.f & 1.3\\ 
\hline
\end{tabular}
\end{table}
\subsection {Physical interpretation of the event}
We carry out photometry of NMS-E1 during its maximum brightness phase using
DAOPHOT (Stetson 1987). The candidate reached its peak brightness in our
observations on December 18, 1999 having magnitudes
\[
R \sim 20.1; ~~~~ I \sim 18.8 ; ~~~~ (R-I) \sim 1.3
\]
There can be an error of $\sim$ 0.2 mag in the colour determination.   
Assuming a distance modulus of $(m-M)_0 = 24.49$ and extinction of
$A_R = 0.63$, $A_I = 0.47$ towards our observed direction of M31 (Joshi et
al. 2003), we estimate
\[
M_R = -5.0; ~~~~ M_I = -6.1; ~~~~ (R-I)_0 = 1.1
\]
\begin{figure}[h]
\centering
\includegraphics[width=9.0cm, height=9.0cm]{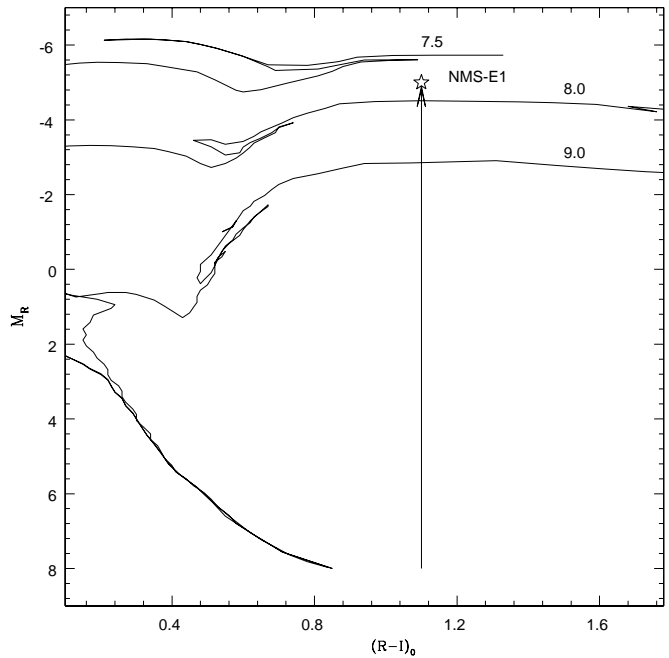}
\caption{The position of NMS-E1 at its maximum brightness in the H-R diagram.
The solar metallicity isochrones by Bertelli et al.~(1994) for log $A$ = 7.5,
8.0 and 9.0 are also plotted. A vertical line is drawn to show the
magnification direction.}
\end{figure}
The position of NMS-E1 in the H-R diagram is shown in Fig.~12. We can see that
NMS-E1 is a very bright red star at its peak brightness which suggests that
the source must lie either on the giant branch or it must be a M1-type
main sequence (MS) star. If it is a MS star, its colour corresponds to an
absolute magnitude of $M_R \sim 8$ mag (Cassisi et al.~1998) hence needs a
very high magnification ($A_{max} > 10^{5}$) to reach a brightness
of $M_R > -5$ mag. Since NMS-E1 is a long time scale event, the latter
possibility is less likely. If it is a red giant star then we should be able
to detect it in the Hubble Space Telescope (HST) images as HST is capable of
detecting giant branch stars in M31. Unfortunately, there are no observations
available for this region in the HST archive. Although NMS-E1 is a long time
scale event and red in colour, it is very unlikely that it is a Mira or any
other long period variable as this event has been monitored for more than 800
days. This is one of the few pixel lensing events reported so far that has
been tracked for such a long duration that allowed us to rule out any
possibility of it being a variable star.

Microlensing events that occur due to the objects in the halo of M31 generally
have a longer time scale than the stars in the bulge or disk of M31 due to the
observer-lens-source geometry (Baltz et al.~2003). The high value of $t_{1/2}$
of NMS-E1 as well as its location on the far side of the M31 disk indicate
that the event might be due to halo lensing. The MEGA collaboration has also
reported two long time scale microlensing candidates in the disk of M31 using
the differential technique i.e. ML-10 and ML-12 with the durations of $t_{1/2}
\sim 47$ and 133 days respectively (de Jong et al.~2003). The candidate ML-10
lies in our observed field at an angular distance of $\sim 6^{'}.6$ from the
position of NMS-E1 but shows a stable light curve in our data since we started
observations when this event had almost merged with the background noise.

\section {Discussion and conclusions}
The main difficulty with the pixel lensing search, in contrast to the
conventional microlensing search, is that one cannot usually be able to measure
the Einstein time $t_E$ of the event from the observed pixel light curve
which carries most of the physical information on the MACHO population. This is
because the flux of a normally unresolved star in the absence of the lensing is
unknown. There is however one event, PA-99-N1 reported by the POINT-AGAPE
collaboration (Auri\'{e}re et al.~2001, Paulin-Henrisksson et al.~2003), where
an unlensed star has been identified in HST archival images. We lack such
identification in the present study. Since we could not identify
the source star, it is not possible for us to determine $t_E$ and
hence the mass of the lens. Nevertheless, by proper modeling of the stellar
populations which either can be source stars or lens objects, we can estimate
various parameters like the timescale of microlensing events, their rate of
occurrence or spatial distribution (de Jong et al.~2003). However, we need
to detect a reasonable sample of genuine microlensing events along with prior
knowledge of the detection efficiency before making any conclusive estimates.

We have reported our first microlensing candidate detected with the pixel
technique on the far side of the galactic disk of M31. The peak brightness
of the candidate $R_{max} \sim 20.1$ mag coupled with its colour $(R-I)_0
\sim 1.1$ and half intensity duration of $t_{1/2} \sim 59$ days lends support
to it being a red giant source while it being main sequence star is highly
improbable. The half intensity duration along with its location in M31 suggest
that the event NMS-E1 might be due to halo lensing. Since we still have not
estimated the detection efficiency, the likelihood of the event and results of
Monte Carlo simulations will be discussed in a future publication. The
variable stars of M31 have also been the subject of numerous studies e.g. star
formation history, stellar population etc.; we are trying to compile a
catalogue of pixel variables in our target field.

Here we presented the results of the pilot observational campaign carried out
towards M31. We have been continuing our observations of the same field
using the new 2-m Himalayan Chandra Telescope (HCT), Hanle, India and we expect
to find a larger number of events which may enlighten us on the dark matter
problem in the galactic halos of M31 and our own galaxy.

~

{\noindent \it Acknowledgments ~}
We would like to thank B. Paczy\'{n}ski for his constructive comments which
improved this paper. It is a great pleasure to thank Yannick Giraud-H\'{e}raud
and Jean Kaplan for their kind support in initiating the project and helping
us to get familiar with the pixel technique. DN is grateful to Japan Society
for the Promotion of Science for an Invitation Fellowship. This study is a
part of the project 2404-3 supported by the Indo-French center for the
Promotion of Advanced Research, New Delhi.

\end{document}